# Big-Data-Driven Materials Science and its FAIR Data Infrastructure

Claudia Draxl[(1, 2)] and Matthias Scheffler[(2, 1)]

1 ) Physics Department and IRIS Adlershof, Humboldt-Universität zu Berlin,
Zum Großen Windkanal 6, 12489 Berlin, Germany
Email: claudia.draxl@physik.hu-berlin.de

2 ) Fritz-Haber-Institut der Max-Planck-Gesellschaft,
Faradayweg 4-6, 14195 Berlin, Germany
Email: scheffler@fhi-berlin.mpg.de

Abstract

This chapter addresses the forth paradigm of materials research – big-data driven materials science. Its concepts and state-of-the-art are described, and its challenges and chances are discussed. For furthering the field, Open Data and an all-embracing sharing, an efficient data infrastructure, and the rich ecosystem of computer codes used in the community are of critical importance. For shaping this forth paradigm and contributing to the development or discovery of improved and novel materials, data must be what is now called FAIR - *Findable*, *Accessible*, *Interoperable* and *Re-purposable/Re-usable*. This sets the stage for advances of methods from artificial intelligence that operate on large data sets to find trends and patterns that cannot be obtained from individual calculations and not even directly from high-throughput studies. Recent progress is reviewed and demonstrated, and the chapter is concluded by a forward-looking perspective, addressing important not yet solved challenges.

## 1.    Introduction

Materials science is entering an era where the growth of data from experiments and computations is expanding beyond a level that can be handled by established methods. The so-called "4 V challenge" – concerning *Volume* (the amount of data), *Variety* (the heterogeneity of form and meaning of data), *Velocity* (the rate at which data may change or new data arrive), and *Veracity* (the uncertainty of data quality) is clearly becoming eminent. Most importantly, however, is that big data of materials science offer novel, extraordinary, and expansive opportunities for achieving scientific knowledge and insight. These opportunities require new research concepts and lines of thought. While this chapter focusses on computational materials science, we emphasize that what is described here applies to experimental data as well.

Today's common approach in computational materials science is to publish results as focused research studies, reporting only those few data that are directly relevant for the respective topic. Thus, even from very extensive computations (expending thousands or millions of CPU core hours) very few results are shared with the community. Most data, in particular when they were not deemed of immediate relevance, are kept private or even thrown away. Since a few years, the community of computational materials science and engineering is undergoing "a change in scientific culture", and has started the extensive sharing of data of this field. Sharing of *all* data, i.e., the full input and output files of computations, implies that calculations don't need to be repeated again and again, and the field will have access to big data which can be used in a totally new research manner, i.e. by artificial-intelligence methods (Draxl and Scheffler 2019 and references therein,). As will be elaborated in the next sections, one can find structure and patterns in big data, gaining new insight that cannot be obtained by studying small data sets[1], and in this way even allegedly inaccurate data can get value. A popular example from daily life of the impact of big-data analytics is the tracking of the movement of mobile phones which provides instantaneous information on traffic flow and jam. Another example is the local information of google searches for *flu symptoms and medicine* which reflect the spreading of a flue wave. The latter example also illustrates the danger of data analytics, as the "google flue trend" worked well for the first few years but then, in 2012, some aspects became unreliable. Reasons may be, among others, anxiety-stimulated searches caused by reports in public and social media and changes in google search technology, e.g. recommending searches based on what others have searched (Lazer et al. 2014). This example also illustrates that big data should not be viewed as substitute but as complement to traditional analysis. Specifically, we note that even if the amount of data may be big, the independent information may be small when data are correlated, or data may be even irrelevant or misleading for the purpose of interest. If such aspects are not considered properly, a statistical analysis will be flawed.

Overcoming the "silo mentality" of computational materials science and the development and implementation of concepts for an extensive data sharing was initiated and achieved by the NOMAD Center of Excellence (NOMAD, Draxl and Scheffler 2019), considering all aspects of what is now called the FAIR data principles (Wilkinson et al. 2016):[2] Data are *Findable* for anyone interested; they are stored in a way that make them easily *Accessible*; their representation follows accepted standards (Ghiringhelli et al. 2016, 2017a), and all specifications are open

---

[1] Let us note that the opposite is true as well, i.e. small data sets can offer information that is hard if not impossible to extract from big data.

[2] The concept of the NOMAD Repository and Archive (NOMAD) was developed in 2014, independently and parallel to the "FAIR Guiding Principles" (Wilkinson et al. 2016). Interestingly, the essence is practically identical. However, the accessibility of data in NOMAD goes further than meant in the FAIR Guiding Principles, as for searching and even downloading data from NOMAD, users don't even need to register.

– hence data are *Interoperable*. All of this enables the data to be used for research questions that could be different from their original purpose; hence data are *Re-purposable*.[3] Obviously, FAIR data also become usable for artificial intelligence tools.

The chapter is structured as follows. In section 2, we briefly summarize the history of the four research paradigms of materials science, with particular focus on the fourth one, "big-data-driven materials science". Section 3 then stresses the importance of an extensive data sharing for the advancements of science and engineering. Section 4 addresses artificial-intelligence concepts for materials-science data with some specific examples. Finally, in section 5, we give an extensive outlook on the developments and open questions of big-data driven materials science.

## 2. Development of the four research paradigms of material sciences

The historical evolution of methods and concepts of materials science are sketched in Figure 1. We recall that experimental research dates back to the Stone Age, and the basic techniques of metallurgy were developed in the Copper and Bronze Ages which started in the late 6th millennium BCE. The control of fire prompted a major experimental breakthrough. Towards the end of the 16th century, scientists began to describe physical relationships through equations. Well-known names from the early days are Tycho Brahe (1546-1601), Tomas Harriot (ca.1560-1621), Galileo Galilei (1564-1642), Johannes Kepler (1571-1630), Isaac Newton (1643-1727), and Gottfried Wilhelm Leibniz (1646-1716). The latter two also developed the concept of the mathematical differential and derivatives. Thus, analytical equations became the central instrument of theoretical physics. Second from the left in Fig. 1, this new way of thinking – the 2. paradigm – is symbolized by the Schrödinger equation. Needless to say, the first paradigm, the empirical and experimental sciences, did not become obsolete, but theoretical physics represents an important complementary methodology.

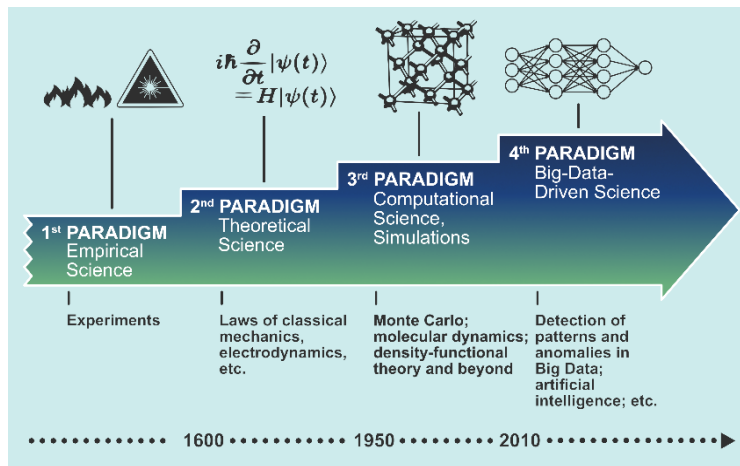

*Figure 1*: Development of research paradigms of materials science and engineering.

---

[3] The NOMAD CoE uses the term re-purposable, while in the FAIR concept it was termed re-usable. The former makes explicit that data can not only be reused but even used for a different purpose.

Since the 1950s electronic-structure theory was advanced for materials by J. C. Slater (e.g. Slater 1937, 1953, 1965, 1967, Slater and Johnson 1972), the Monte-Carlo method was introduced by (Metropolis et al. 1953), and (Alder and Wainwright 1958, 1962, 1970) and (Rahman 1964) introduced molecular dynamics. (Hohenberg and Kohn 1964) and (Kohn and Sham 1965) laid the foundation of density-functional theory (DFT)[4] in the mid nineteen-sixties. All these developments enabled computer-based studies and analysis of thermodynamics and statistical mechanics on the one hand, and of quantum-mechanical properties of solids and liquids on the other hand. They define the beginning of computational materials science, what is nowadays considered the 3rd paradigm. Herewith "computer experiments" were introduced, i.e. simulations, whose results are often treated and analyzed analogous to those of experimental studies. Initially developed independently, the fields of electronic-structure theory and statistical mechanics and thermodynamics are now growing together (Reuter et al. 2005 and references therein). Likewise, in parallel to DFT, many-body techniques, based on Green functions were developed (Hedin 1965), which are now synergistically interleaved with DFT to form the forefront of electronic-structure theory, including excitations.

Today, big data and artificial intelligence revolutionize many areas of our life, and materials science is no exception (Gray 2007, Agrawal and Choudhary 2016). Jim Gray had probably first discussed this 4th paradigm arguing explicitly that big data reveal correlations and dependencies that cannot be seen when studying small data sets. Let us generalize the "big data" concept by noting the complexity of materials science (and others sciences as well): The number of potential but initially unknown descriptive parameters that characterize or identify the properties and functions of interest may be very big. Thus, the diversity and complexity of mechanisms represents a big-data issue in materials science as well. A further important difference to the second paradigm is that we accept that many materials properties, i.e. patterns and correlations in big data, cannot be described in terms of a closed mathematical formulation, as they are governed by several, intermingled theoretical concepts and multilevel, intricate processes. As a consequence, such patterns represent and advance knowledge but they do not necessarily reflect understanding.

---

[4] See the Chapter by (Li and Burke 2018)

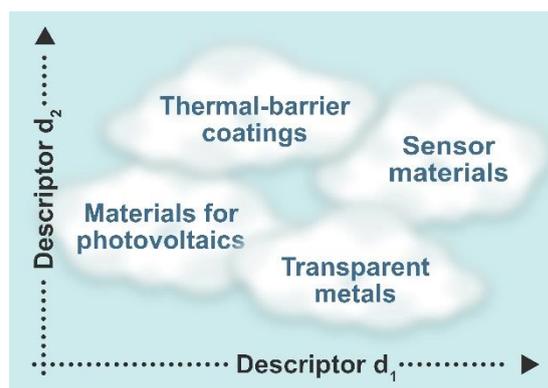

**Figure 2.** Big data contain correlations and structure that are not visible in small data sets. Finding descriptors that determine a specific property or function of a material is a crucial challenge. Once this is in place, we will be able to machine learn the data and eventually draw maps of materials. In difference to the sketch, these maps will be typically of higher dimension than just two.

What is our vision for exploiting the fourth research paradigm of our field? Figure 2 provides a schematic view on it: The space of different chemical and structural materials is practically infinite. However, when asking e.g. for high-performance thermal-barrier coatings, there are just a few suitable materials. Thus, in terms of functional materials the materials space is sparsely populated as most of the already existing or in the future synthesized materials are irrelevant. Hence, the relevant data are a statistically exceptional minority, and if this situation is ignored, a statistical analysis that assigns the same importance to all data may well be problematic. Finding regions or patterns that correspond to materials with superior functional performance requires the identification of appropriate descriptors, noted as $d_1$ and $d_2$ in Fig. 2. Obviously, in general, the dimensionality will likely be higher than just two. At this point, Fig. 2 is just a sketch, and as for most properties, the appropriate descriptors are unknown, the scientific challenge is to find appropriate ones. Let us emphasize, as the actually relevant data is not big enough (in most cases of interest in materials science), it is important if not crucial to use as much domain knowledge as possible. Compressed sensing, subgroup discovery, machine learning, and other methods of artificial intelligence are able to identify these descriptors and patterns, and we will address these methods in section 3 below.

We close this section by noting the radically decreasing time scales of new developments and paradigms: 6[th] millennium BCE, 1600, 1950, 2010. Thus, a next research paradigm may be just ahead of us.

### 3. Extensive data sharing – why and how?

Data are a crucial raw material of this century. Our community is producing materials data by CPU-intensive calculations since many years. Typically, the results are stored on PCs, workstations, or local computers, and most of these data are not used and often even thrown away, though the information content could well be significant. The field is now slowly changing its scientific culture towards

*Open Science* and *Open Data*, and there are many reasons for doing so. Open access of data implies that data can be used by anyone, not just by the experts who develop or run advanced codes. If data were openly available (and well described), many more people would work with the data, e.g. computer scientists, applied mathematicians, analytic condensed matter scientists, and more. We will be surprised what people will do with data when they are made available, probably using tools that the present computational materials science community does not even know.

Let us mention one example, a "crowd sourcing" data-analytics competition at the internet platform Kaggle (Kaggle/NOMAD2018). This competition addressed the search for novel transparent and semiconducting materials using a dataset of $(Al_xGa_yIn_z)_2O_3$ compounds (with $x+y+z=1$). The aim of this challenge was to identify the best machine-learning model for the prediction of two key physical properties that are relevant for optoelectronic applications: the electronic band-gap energy and the crystal-formation energy. These target properties were provided for 2,400 systems covering 7 different crystal space groups and unit cells ranging from 10 to 80 atoms. 600 additional systems made the test set for the competition.

The competition was launched in December 2017, and when it finished, 2 months later, an impressive number of 883 solutions had been submitted by researchers or research teams from around the world, employing different methods. Interestingly, the three top approaches, summarized in a recent publication (Sutton et al. 2019), adopted completely different descriptors and regression models. For example, the winning solution employed a crystal graph representation to convert the crystal structure into features by counting the contiguous sequences of unique atomic sites of various lengths (called *n*-grams), and combined this with kernel ridge regression (KRR). This *n*-grams approach was new for materials science.

To enable the envisioned success of big-data-driven materials science, the field obviously needs a suitable data infrastructure that facilitates efficient collection, data description in terms of metadata, and sharing of data. For the field of computational materials science, this was developed by the NOMAD Center of Excellence (NOMAD) which also instigated comprehensive data sharing. The synergetic relationship with other major key data bases, in particular AFLOW, Materials Project, OQMD made it the biggest repository in the field. Uploaded data are tagged by a persistent identifier (PID), and users can request a DOI (digital object identifier) to make data citable. Data downloading does not require any registrations and refers to the Creative Commons Attribution 3.0 License. On a more formal basis, and in parallel to NOMAD, the proper way of collecting data was suggested as the *FAIR Guiding Principle*s (Wilkinson et al. 2016).[2,4]

So what is behind FAIR? What does it mean for computational materials science?

**The *F*** stands for ***findable***. Making research data open and keeping them for at least ten years is now requested by many research organizations. Seen from a practical point of view, it is also useful to avoid doubling of work and thus save human and computational resources and energy. Since individual researchers create their data on various platforms – from workstations to compute clusters to high-performance computing (HPC) centers –, it is often impossible to find a student's or postdoc's data, some time after s/he has left the research group. Besides matters of organization, issues may be related to automatic erasure of data in HPC centers, missing permissions on local machines, data protection, and alike. Clearly, making data findable requires a proper data infrastructure, including documentation, metadata, search engines, and hardware. This is one of the reasons why the NOMAD Repository and its metadata[5] was established.

**The *A*** stands for *accessible*. Accessibility in materials science has different facets. First, we should not forget about the proper hardware that allows for swift access to data. Second, we need to provide application programming interfaces (APIs). To make data fully accessible requires an important additional step namely the formal description of the data, i.e. its metadata that also consider the metadata interrelations[5]. This connects to the *I* in FAIR.

**The *I*** stands for *interoperable*. Here we need to consider in a first place the extreme heterogeneity of computational materials data. The wider community is using about 40 different, major computer codes (considering electronic-structure, molecular-dynamics, and quantum-chemistry packages for materials) that differ in various aspects of methodology and implementation. Consequently, the necessity arises to make their results comparable, which is a major challenge not only in the sense that they need to be brought to a common format and common units; we also recall that one quantity may be named differently in different (sub-)communities or one and the same expression may have a different meaning in one or the other area. Thus, "dictionaries" are needed to translate between them. Obviously, if one would restrict everything to just one computer program package, translations or conversion are not necessary. However, then a significant part of the available data and, even more importantly, of the community would be missed. In this sense, the NOMAD concept is general and in fact orthogonal to essentially all other data repositories that typically focus on just one computer code (see Draxl and Scheffler 2019).

Still, we need to ask if one can operate upon all available data in a meaningful way. Apart from formats and units, there are more severe restrictions that may hamper such important undertaking. These concern the computational parameters – and consequently the numerical precision – that are used in different calculations. We recall here that for extensive sharing, all open data can be considered valid in the sense that, when the code and code version are provided, output files[6] correspond to the provided input files in a reproducible way. Nevertheless, data have been created for different purposes which may require different levels of convergence in terms of basis-set size and alike.[7] More than that, we may even ask whether different codes aiming at the solution of one and the same problem with "safe" settings give the same results. For the latter, we point to the community effort led by Stefaan Cottenier (Lejaeghere et al. 2016), where the equations of state for 71 elemental solids were calculated with many different *ab initio* electronic-structure packages. Over a time span of a few years, it turned out that upon improvements of codes, basis sets, and in particular pseudopotentials, all codes lead to basically the same answer. In fact, the first thirteen entries in the list[8] differ by an average absolute error of less than 0.5 meV per atom in total energies. Such investigations are extremely helpful and have set the stage towards a culture of benchmarking, which had been established in quantum chemistry for molecules already decades ago. The study by (Lejaeghere et al. 2016) is, however, only the beginning. Clearly, other properties like energy barriers, band gaps, spectra, etc., and systems like surfaces, interfaces and inorganic/organic hybrid materials, etc. will be much less forgiving than total energies of simple bulk solids and will make discrepancies more obvious. Therefore, more investigations along these lines are on the way.

---

[5] The NOMAD CoE has developed extensive metadata, generic as well as those specific to 40 community codes (Meta Info at NOMAD).

[6] Without the full input file and the main output file(s) (NOMAD) does not accept uploads.

While the above comparison (Lejaeghere et al. 2016) could only be made with parameter sets that represent full convergence, calculations performed in daily life, are often far from this optimal case, and are, in fact, often sufficient. This situation obviously leads to the question how to compare and operate on calculations that have been performed with different settings, e.g. in terms of basis sets, meshes for Brillouin zone integrations, and alike. Below, it is shown that this is, in fact, possible (Carbogno et al. 2019).

Let us demonstrate that fully converged results of *complex* materials can be estimated by learning from errors of calculations of *simple* materials (Carbogno et al. 2019). Four different codes have been employed for this study: Two very different all-electron codes and two projector augmented wave/plane-wave codes. These are: exciting (Gulans et al. 2014) and FHI-aims (Blum et al. 2009), and GPAW (Enkovaara et al. 2010), and VASP (Kresse and Furthmüller 1996). Sources for discrepancies of different calculations in publications and/or the various data repositories are incomplete basis sets, approximate treatment of the **k**-space integration, the use of pseudopotentials, and more. Since incomplete basis sets are, indeed, the most severe issue, Fig. 3 shows how the total energies for fixed geometries change as a function of basis set quality for the 71 elemental solids adopted from (Lejaeghere et al. 2016. The red symbols in Fig. 3 mark the materials exhibiting the largest error and the black symbol refer to the average taken over all materials. The error, $\Delta E$, is defined for each material with respect to the fully converged value obtained with settings as or even more precise than the ones used in Lejaeghere et al. 2016. In all cases, the error decreases systematically from the order of 1eV for small basis sets down to meV precision.[9] Based on these errors of the elemental solids, Carbogno and coworkers (Carbogno et al. 2019) estimated the errors arising from the basis-set incompleteness in multi-component systems, utilizing a simple analytic model, i.e., by linearly combining the respective errors of the constituents (elemental solids) obtained with the same settings:

$$\overline{\Delta E_{tot}} = \frac{1}{N} \sum_N N_i \Delta E_{tot}^i . \qquad (1)$$

---

[7] For example, the determination of the atomic geometry may need less stringent parameters than details in the electronic band structure. We also note that what was regarded "converged" a few years ago, may not be considered precise enough today. This should not devalue older calculations (see also the discussion of Figs. 3 and 4).

[8] At https://molmod.ugent.be/deltacodesdft, one has to choose a reference where obviously an all-electron code is the natural choice. In fact, the precision of the all-electron packages WIEN2k, exciting, and FHI-aims are practically identical, and these codes are leading the whole list.

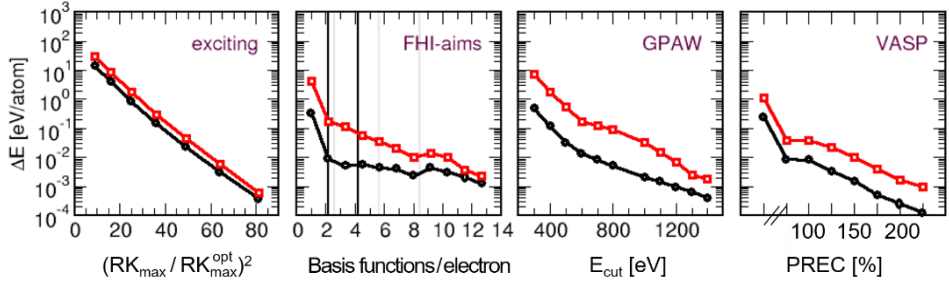

*Figure 3:* Convergence of total energies of 71 elemental solids with increasing basis-set quality for four different electronic-structure codes. Red symbols indicate materials with maximal error, black symbols display averages over all 71 materials. All values are taken with respect to the respective fully converged results. $RK_{max}$ is the LAPW-specific parameter defining the basis set size, $E_{cut}$ the GPAW cutoff energy, PREC is the VASP-specific parameter for basis set quality. The black lines for FHI-aims indicate light (left) and tight (right) settings, the grey lines Tier1, Tier2, and Tier3 (from left to right). Note that the errors are displayed at a logarithmic scale.

Here $N_i$ is the number of atoms of species *i* present in the compound, and $\Delta E_{tot}^i$ is the error of this species in the respective elemental solid. This model is applied to 63 binary solids that were chosen such to optimally cover the chemical space (one for each element with atomic number up to 71, without noble gases). This approach is validated by comparing the estimated errors to those of corresponding DFT calculations for these binaries, as depicted in Fig. 4. The authors of this work find a clear correlation between predicted and actual errors for all four codes (with the details depending on peculiarities of the specific basis set). For more in-depth analysis, also including relative energies and ternary alloys and an extensive discussion about the role of the employed method/basis set, we refer the reader to (Carbogno et al. 2019). In essence, the conclusion can be drawn that even when not having fully converged values for complex materials in hand, one can estimate the energetics for the fully converged case. This study sets the stage for data coming from different sources fulfilling an important condition for the *I*.

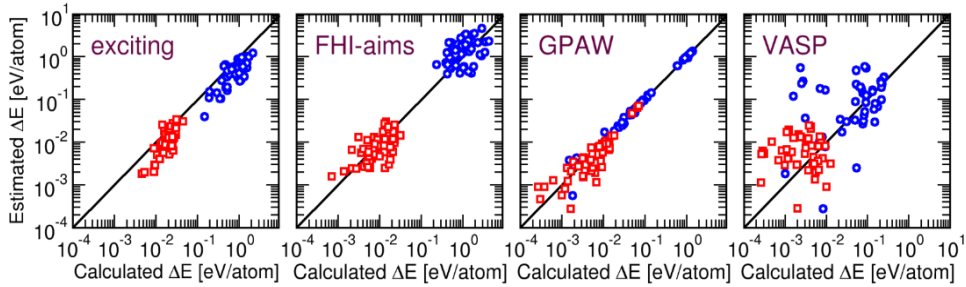

Figure 4: Predicted versus actual errors in total energy for 63 binary alloys, obtained from the errors in elemental solids for four different electronic-structure codes. Blue data are from calculations performed at low-precision settings, and red data are from "standard quality" settings. Note that they are close to the straight black line. For details see (Carbogno et al. 2019).

---

[9] The fact that not all codes show a monotonous behavior for small basis set sizes has to do with specifics of the basis set. Here we refer the reader to (Carbogno et al. 2019).

**The *R*** stands for *re-usable*. In fact, we prefer the term *re-purposable* that gives a better impression about what is meant in the materials-science context. It means that we can use data that has been created for some specific scientific question, in a different connection. Indeed, one and the same material can be considered for various applications. So why should a researcher working on one aspect not allow another researcher to use the same data for focusing on a different aspect? Let us illustrate this with the example of $TiO_2$ which is an important support material for heterogeneous catalysis. The detailed results are not only useful for researchers working in this area, but also in a different context. For example, for photovoltaics $TiO_2$ is a component of dye-sensitized solar cells. And, as another example we note that $TiO_2$ is used as pigment in paints and in cosmetic products e.g. sunscreens.

It is generally agreed that research results obtained in academia should be published. In view of what was discussed above, it should be a duty to publish all the results, i.e. making also the complete data underlying a publication available. This has been said by many people, funding agencies, politicians, and research journals. Indeed, a few research journals have started to request that all details are uploaded at a certified repository. Obviously, as mentioned above, data must be connected to established metadata and to workflows[10] of their production. However, to fully apply such concepts, the necessary data infrastructure hardly exists so far. Let us cite from a recent Nature Editorial "Empty rhetoric over data sharing slows science" (Nature editorial 2017): "Everyone agrees that there are good reasons for having open data. It speeds research, allowing others to build promptly on results. It improves replicability. It enables scientists to test whether claims in a paper truly reflect the whole data set. It helps them to find incorrect data. And it improves the attribution of credit to the data's originators. But who will pay? And who will host?" – and further "Governments, funders and scientific communities must move beyond lip-service and commit to data-sharing practices and platforms." For computational materials science though, NOMAD had already changed the culture and implemented an appropriate platform (Draxl et al. 2017, Draxl and Scheffler 2018, 2019).

---

[10] In technical terms "workflow" refers to the sequence and full description of operations for creating the input file (or the sample in experiment) and performing the actual calculations (or the experiment). In computational materials science, many input and output files provide the information about the various steps of the workflow. Important workflow frameworks that allow to automatically steer, analyze, and/or manage electronic-structure theory calculation are ASE (atomic simulation environment) (Larsen et al. 2017), Fireworks (Jain et al. 2015), AFLOW (Calderon et al. 2015), and Aiida (Pizzi et al. 2016).

**4 Artificial intelligence concepts for materials science data**

We are using "artificial intelligence (AI)" as umbrella term of computational methods that "learn from experience". As already mentioned, for materials science, the complexity of the actuating mechanisms is big but the number of relevant data is typically on the lower side. Zhang and Ling (2018) recently addressed this issue in terms of the degree of freedom of the model and the prediction precision. In a similar sense we argue that a proper analysis of data needs to consider at least some aspects of the causality that drives the correlations of interest, i.e. one needs to include domain knowledge in the learning process in order to achieve a trustworthy description, interpretability, and possibly even deeper understanding of the cause behind the structure or patterns in the data. Specifically, we note that from the practically infinite number of possible materials only 10 or 100 may be relevant for a certain purpose. Standard machine learning tools address a description of the crowd or the majority and therefore optimize the root-mean-square error or alike and also introduce a regularization to avoid overfitting and/or to achieve a smooth description. As a result, it is likely that statistically exceptional data are suppressed. But only these carry the information we are interested in. The lower the employed domain knowledge is, the larger is the amount of data that is needed for the learning process, and it may happen that data are fitted but predictions and even interpolations are not reliable. AI is a wide and interdisciplinary field, and machine learning (learning from data) and compressed sensing (originating from signal compression; primarily aiming at modeling properties or functions in terms of low dimensional descriptors) are important subdomains.

As noted above (see the discussion of Fig. 2), big data may reveal correlations (structure and patters) if and only if the data are arranged in a proper way, e.g. represented by appropriate descriptors. These correlations can be found by AI but the identification of such correlations does not necessarily go along with deeper insight or understanding. To some extent we like to argue, that the wish for insight and understanding is often overrated. This is well documented by the Periodic Table of the Elements that may arguably be considered as one of the most impacting achievements for chemistry, condensed matter physics, engineering, and biophysics. When Mendeleev published his table in 1871, based on knowledge of 63 atoms (their weights and chemical behavior), there was no understanding of the deeper cause behind the structure of the table (Scerri 2008; Pyykkö 2012). Still, the table predicted the existence of at that time unknown atoms, and even their properties where described. However, the underlying causality, i.e. that the different rows reflect the different number of nodes of the radial wave functions of the outer valence electrons, and that the different columns refer to the number of valence electrons, was unknown when the table was created. It only was understood about 50 years later, when the shell structure of electrons in atoms was described by quantum mechanics.

Thus, identifying correlations, structures, and patterns in big data is an important step by its own. When the relationship between a property of interest, *P,* and a set of useful descriptive parameters (the descriptors $d_1$, $d_2$, ... – also called representation) is known, graphs as in Fig. 2 or approximate equations can be obtained for the relationship $P(d_1, d_2, ...)$. For the example of the Periodic Table, the descriptors are the row and column numbers. Obviously, as the number of possible materials is practically infinite, building a map as in Fig. 2 is a highly demanding task, of much higher complexity than building the Periodic Table of the Elements.

How to find the descriptors for materials properties? The direct and complete descriptor for a quantum-mechanical materials problem is given by the position of all atoms, the nuclear numbers, and the total number of electrons: $\{\boldsymbol{R}_I, Z_I\}$, $N^e$. These descriptive parameters fully identify the many-body Hamiltonian but learning *the properties* that result from a given Hamiltonian is a very demanding goal. Thus, the amount of data needed for training (fitting) a materials property or function directly in terms of $\{\boldsymbol{R}_I, Z_I\}$, $N^e$ is typically unrealistically high. Instead, for choosing proper descriptors, that relates to the actuating mechanism of interest, we distinguish three concepts: a) the descriptor may be selected out of a huge, systematically created pool of candidates; b) the descriptor may be built in the machine-learning step in an abstract manner; and c) one may just use a (known) descriptor assuming that with many data the actual choice may be not so important. Concept a) will be discussed below when we describe compressed sensing and subgroup discovery. Concept b) is realized in neural-network approaches which, in the learning step, minimizes an objective function that quantifies the difference between the predicted and the correct (known) data. Through this minimization, the weights (i.e. parameters) of the neural network are optimized (Hinton 2006, Hinton et al. 2006), and in this way, the network learns the descriptors. Doren and coworkers and Lorenz et al. (Blank et al. 1995, Lorenz et al. 2004, 2006) have shown early examples of representing potential-energy surfaces of materials by neural networks. Hellström and Behler describe recent advances in their chapter (Hellström and Behler 2018). Concept c) is probably the most widely used approach. Example descriptors that encode the chemical and geometrical information are Coulomb matrices (Rupp et al. 2012, Hansen et al 2013), scattering transforms (Hirn et al. 2015), diffraction patterns (Ziletti et al. 2018), bags of bonds (Hansen et al. 2015), many-body tensor representation (Huo and Rupp (2017), smooth overlap of atomic positions (SOAP) (Bartók et al. 2010, 2013), and several symmetry-invariant transformations of atomic coordinates (Seko et al. 2017, Schütt et al. 2014, Faber et al. 2015). This concept is nicely described in the chapter by Ceriotti, Willatt, and Csányi (Ceriotti et al. 2018).

A systematic understanding of the suitability of various machine-learning (ML) models and thorough benchmarking studies are still lacking in materials science. It was only recently addressed in terms of a public data-analytics competition that was hosted on the internet platform Kaggle using a data set of 3,000 $(Al_xGa_yIn_z)_2O_3$ compounds ($x+y+z=1$), already mentioned in section 3 above (see Sutton et al. 2019).

Some caution may be appropriate. In general, an observed correlation will have a causal reason – provided that it is supported by a sufficiently large data set (Pearl 2009). Thus, a correlation that is described by the function $P(d_1, d_2, ...)$ may reflect that $(d_1, d_2, ...) = \boldsymbol{d}$ are the actuators: $\boldsymbol{d} \rightarrow P$. However, it could well be that the reverse is true: $P \rightarrow \boldsymbol{d}$. Thirdly, it is possible that there is an "external master", $M$, who controls both $\boldsymbol{d}$ and $P$, with no direct relationship between $\boldsymbol{d}$ and $P$. And fourthly, the data may be selected with a significant bias of the researcher or research community. We fear that this may be happening much more frequently than realized. But then the observed correlation may just reflect this bias. All this needs to be kept in mind when tools of artificial intelligence are applied to big (or not so big) data and when we ask for interpretability or even causality.

Let us add another warning about big data of materials science. The number of possible materials is practically infinite, and we like to identify new materials that have better performance or functionality than the materials that are used today. Clearly, the amount of available data in materials science is getting big though from the few (about 250,000) inorganic materials that have been synthesized up to now, we often just know the atomic structure and hardly their electronic, elastic or other properties. Getting more and more data, however, does not imply that all the data are relevant for all properties of interest. Materials science shows a high diversity, i.e. it is ruled by a significant number of different properties and mechanisms, and experience seems to show that at the end, the number of materials that are good for a certain group of functions is very small. For example, if we ask for a highly transparent materials with excellent heat conductivity and scratch resistance there is probably nothing better than and nothing even close to diamond. Another example is the recent study by (Singh et al. 2019) who studied 68,860 candidate materials for the photocatalytic reduction of $CO_2$. Only 52 of them turned out to be potentially relevant. In general, it is likely that in the few "high-performance materials" a mechanism is active (or inactive) that is not relevant (or dominant) in the majority of materials. Thus, the amount of available data may be big but the number of *relevant* data, i.e. data that contain the needed information, is small. In simple words, in materials science and engineering, we are often looking for "needles in a hay stack", i.e. for very few systems that are statistically exceptional, in some ways. Fitting all data (i.e. the hay) with a single, global model may average away the specialties of the minority (i.e. the needles). Thus, we need methods that identify the relevant, possibly small, statistically-exceptional subgroups in the large amount of data right away.

Let us sketch this challenge for kernel methods of machine-learning approaches which presently play a significant role in the field. The property of interest is written as a sum over a large, appropriate subset of all known data $j$

$$P(\boldsymbol{d}) = \sum_{j=1}^{N} c_j K(\boldsymbol{d}, \boldsymbol{d}_j) \quad . \tag{2}$$

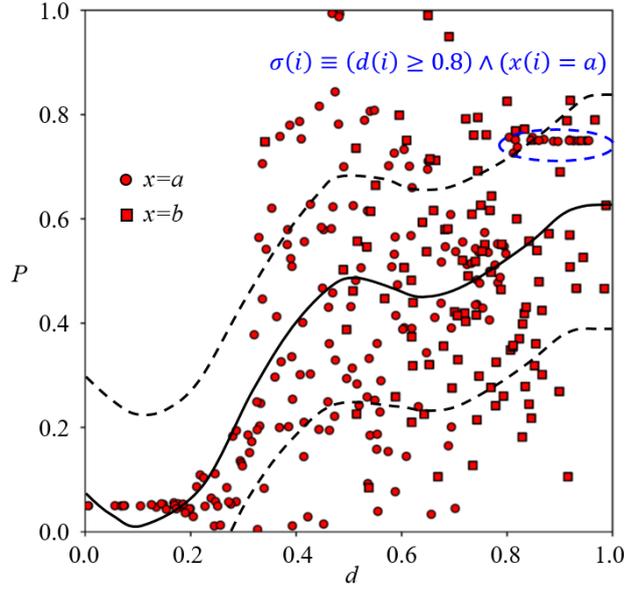

Figure 5: Sketch for a machine learning method, i.e. fit of 1,000 data points (full line) and the confidence interval, which contains 69% of all the data using Gaussian process regression. Also noted is a subgroup (circled by a blue dashed line) that is statistically exceptional but not treated correctly in the global machine-learned description. The selector equation is noted in blue (Boley 2017). For details see the subsection on subgroup discovery.

There are many options for the kernel *K*. A most popular and very successful choice is the Gaussian kernel

$$K(\boldsymbol{d}, \boldsymbol{d}_j) = \exp(-(\boldsymbol{d}-\boldsymbol{d}_j)^2/2\sigma_j^2) \quad . \qquad (3)$$

Fitting a set of say *N*=100,000 known data is achieved by determining 100,000 coefficients by minimizing

$$Min\,\{\sum_{i=1}^{N}(\hat{P}_i - P(\boldsymbol{d}_i))^2 - \lambda ||m||\}\quad . \qquad (4)$$

Here $\{\hat{P}_i\}$ are the actually known data that should be fitted, and we also introduced a regularization which prevents overfitting and creates a result that does not go exactly trough the known data points but is smooth. This regularization is noted as "norm of the applied model, *m*". Figure 5 gives an example of such fitting/machine-learning approach. Let us, at first, ignore that these are two types of data (noted as squares and circles): Obviously, fitting *N* data points with a function that contains *N* free parameters must work, but the regularization creates some uncertainty (a smooths curve), and the interpretability of the many determined coefficients is typically lacking. Figure 5 also shows a subgroup (statistically exceptional data), together with its selector. These data are not described well by the employed kernel approach. Unfortunately, typically we don't know which data are relevant and which are not. Otherwise a weighted sampling could be imposed (Chawla et al. 2002 and references therein). The example reveals that the general statement "more data provide a better description" does typically not apply to ML for materials science as it may just mean: Add more irrelevant information (more hay) to the information pool (the hay stack). This will not help to find the needles. Alternatively, could we turn this around? Can we attempt to fit the hay and then consider, the few materials that are distinguished by a high fitting error as an interesting subgroup that contains the needles? The difficulty here is that materials are very heterogeneous, and this heterogeneity is not just restricted to the direct hay-needle comparison. Also the "hay" is heterogeneous and a high fitting error could also result from over- or underfitting and is not necessarily correlated with the target properties.

Nevertheless, whenever we attempt a global description, machine learning is a great tool. The chapter by (Huang et al. 2018) gives an excellent description, and the above mentioned work on metal oxides (Sutton et al. 2019) is a good example.

Two interpretability-driven approaches have recently been adopted by materials science. These are *subgroup discovery* on the one hand and *compressed sensing* on the other. Let us introduce them briefly.

*Subgroup discovery*

As noted above, a global model addressing the quantitative description of the entire data set may be difficult to interpret. For many requirements in materials research and development, local models that identify and describe subgroups would be advantageous. For illustration (see Fig. 6), a globally optimal regression model could predict a negative relationship between the data (Fig. 6 left). However, a subgroup discovery analysis may reveal that there are two statistically exceptional data groups (indicated by blue and red color in the right part of the figure). Thus the relationship in the data set does not have a negative slope (the global model) but positive slopes (the two subgroups). As a physical example, the transition metals of the Periodic Table are a subgroup, and the actinides, lanthanides, and halogens are other subgroups. Thus, identification of subgroups is useful to gain an understanding of similarities and differences between systems.

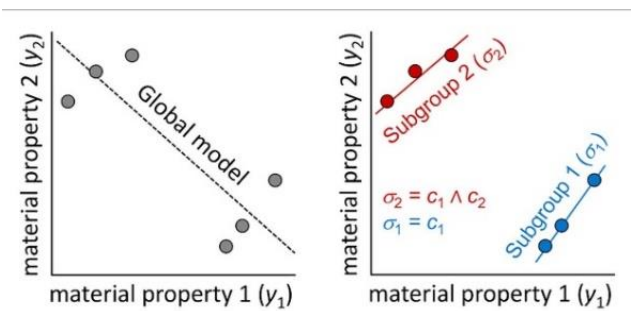

**Figure 6:** Left, data points and a global regression machine-learning model predicting a negative relationship between material properties $y_1$ and $y_2$. Right, subgroup discovery identifies statistically exceptional regions marked as red and blue, and machine learning of these two regions exhibits positive slopes. Subgroup discovery aims at describing such subpopulations by Boolean selector functions ($\sigma_1$ and $\sigma_2$) defined as conjunctions (logical AND, denoted as $\wedge$) of basic selectors (the $c_i$).

The concepts of subgroup discovery (SGD) was introduced in the early 1990s, when the advent of large databases motivated the development of explorative and descriptive analytics tools as an interpretable complement to global modeling (Duivesteijn et al. 2016, Klösgen 1996, Atzmueller 2015, Herrera et al. 2011, Siebes 1995, Wrobel 1997, Friedman and Fisher 1999). Simply speaking, the identification of subgroups is built on 3 components: *i)* The use of a description language for identifying subpopulations within a given pool of data. These are Boolean expressions, e.g. "the ionization potential of atom A minus the ionization potential of atom B is smaller than $X$" where $X$ is a number that may be chosen iteratively. These expressions are called selectors. *ii)* The definition of utility functions that formalize the interestingness (quality) of a subpopulation. This may include requests as "the band gap of the material is in between 1.1 and 1.5 eV" AND "the cohesive energy is larger than 3 eV"; and *iii)* The design of search algorithms to find selectors that describe the subpopulations of interest (Goldsmith et al. 2017). Figure 7 illustrates the idea for a recent study of heterogeneous catalysis: Finding potential catalysts that can transform the greenhouse gas $CO_2$ into useful chemicals or fuels (Mazheika et al. 2019). This study concentrated on metal oxides and realized that a global model (fitting all the data at once) failed to provide an accurate description. However, searching for subgroups by considering several potential indicators for a good catalytic activity and many potential selectors reveals several subgroups that are statistically exceptional. Only one of them (marked in green in Fig. 7) contains the known catalytically active materials. Details can be found in (Mazheika et al. 2019). The study identified a new indicator for the catalytic $CO_2$ activation, and it provided several suggestions for new catalysts.

*Compressed Sensing and the SISSO Approach*

As noted in the discussion of Fig. 2, finding a descriptor (2-dimensional in Fig, 2), i.e. the set of parameters that capture the underlying mechanism of a given materials property or function, is the key, intelligent step toward identification of structure or patterns in (big) data. This central role of the descriptor was only recently addressed explicitly and systematically in the works of Ghiringhelli and coworkers (Ghiringhelli et al. 2015, 2017b; Ouyang et al. 2018). These authors recast the descriptor-identification challenge into a systematic formulation using compressed sensing (CS).

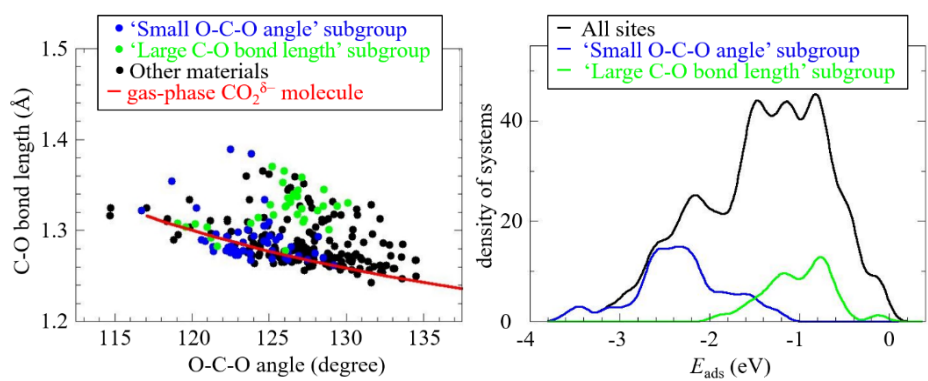

*Figure 7*: Two subgroups of oxide materials in a data set that describes the $CO_2$ adsorption. The blue subgroup is characterized by "small" angles of the O-C-O molecule (the neutral free molecule is linear (180°)). And the green subgroup is characterized by at least one large C-O bond length (the bond lengths of the neutral free molecules are 1.17 Å). Both subgroups reflect a weakening of the bond, but only the green one correlates with a good catalytic activity. Right: Density of systems of the full data set and of the two subgroups as function of adsorption energy.

The CS approach had been originally introduced for efficiently reproducing a high-quality signal from a very small set of observations (Candès and Wakin 2008, Nelson et al. 2013, Candès et al. 2006, Candro et al. 2006, Donoho 2006). Mathematically, this can be sketched as follows. Given a set of data $P_1$, $P_2$, … $P_N$, where $i$=1-$N$ labels different materials (or different conditions), CS finds the sparse solution $c$ of an underdetermined system of linear equations

$$P(\boldsymbol{d}_i) = \sum_{k=1}^{M} \hat{c}_k \, d_{ki} \,. \qquad (5)$$

$\{d_{ki}\} = \boldsymbol{D}$ is called the sensing matrix with the number of rows $k$=1-$M$ significantly bigger than the number of columns, $i$=1-$N$. Thus, the sensing matrix is built from $N$ vectors (the columns), each of length $M$. Material $i$ is characterized by vector $i$, i.e. by $k$=1-$M$ descriptive parameters, $d_{ki}$. Equation (5) corresponds to Eq. (2) when the linear kernel is used. If most elements of the vector $\hat{\boldsymbol{c}}$ were zero, specifically when the number of nonzero elements of $\hat{\boldsymbol{c}}$ is smaller than $N$, the dimensionality of the problem is reduced (Candès et al. 2006; Donoho 2006; Candès and Wakin 2008). In order to achieve this reduction, the coefficients $c_k$ are determined by solving Eq. (4) with the norm ||m|| taken as the $l_0$ norm of $\hat{\boldsymbol{c}}$. The norm zero of a vector is defined as the number of non-zero elements. Thus, the regularization $\lambda \|\hat{\boldsymbol{c}}\|_0$ is a constraint that favors solutions for $\hat{\boldsymbol{c}}$ where most elements of $\hat{\boldsymbol{c}}$ are zero. However, using the norm zero poses a combinatorically extensive problem, and it has been shown that this is (asymptotically) NP hard. As a consequence it has been suggested to approximate the norm zero by the norm $l_1$, and a popular approach to it is LASSO (least absolute shrinkage and selection operator) (Tibshirani 1996). For materials science this has been introduced by Ghiringhelli and coworkers (Ghiringhelli et al. 2015, 2017b).

Thus, the idea is to offer many descriptor candidates and then let the optimization approach (Eq. (4)) find out which of these candidates are best. Since Eq. (5) is linear, it is necessary that the offered descriptor candidates contain the potential nonlinearities. Consequently, different descriptors, i.e. different columns of the sensing matrix, may become correlated. Furthermore, the dimension of the sensing matrix increases rapidly with the number of data points, and as LASSO requires that the matrix is stored, the approach is getting intractable. These problems have been recently tackled by Ouyang and coworkers (Ouyang et al. 2018) by solving the $l_0$ challenge in an iterative approach called SISSO (sure independence screening and sparsifying operator). Interestingly, the mentioned correlations are not causing problems, and the number of candidate descriptors can be increased in SISSO to many billions and even trillions. Initially, from the previously mentioned "basic descriptors" $\{\boldsymbol{R}_I, Z_I\}$, $N^e$ only $Z_I$ derived quantities were used explicitly, e.g. the ionization potentials of the atoms, the electron affinities, and information about the extension of the atomic wave functions. Then, a combination of algebraic/functional operations is recursively performed for extending the space of potential descriptors. The operators set is defined as $+, -, \times, /, \exp, \log, |-|, \sqrt{}, ^{-1}, ^2, ^3$. Details are described in (Ouyang et al. 2018). Clearly, when different structures are considered or different charge states $\{R_I\}$, $N^e$ related features are needed as well.

## 5. Outlook

Computational materials science took off with impressive early work by Moruzzi, Janak, and Williams (Moruzzi et al. 1978) on various properties of metals and by Cohen and coworkers (Yin and Cohen 1982) on the cohesion and phase transition of silicon and germanium[11]. A number of computer codes for solving the Kohn-Sham equations have been developed since then, initially involving approximations like pseudopotentials (removing the core electrons, creating smooth potentials) or introducing touching or slightly overlapping atom-centered spheres in which the potential was sphericalized. During the 1980's significant advancements in the original pseudopotential approach have been made (see the work of Vanderbilt and coworkers: Garrity et al. 2014 and references therein), and all-electron codes have been developed that don't require a shape approximation for the potentials (e.g. Blaha et al. 1990, Gulans et al. 2014, Blum et al. 2009). The work by (Lejaeghere et al. 2016) provides a nice overview of the precision of modern electronic-structure codes for elemental bulk solids, also demonstrating how to involve the community. Clearly, this kind of work is important and needs to be extended to more complicated structures and compositions, defects, surfaces, and interfaces. Work in this direction is underway, as are studies for advanced electronic-structure methods, like e.g. the *GW* approach (van Setten et al. 2015). Furthermore, the field urgently needs benchmarks for the various numerical approximations and for exchange-correlations potentials in order to address also accuracy, not only numerical precision. The MSE (materials science and engineering) project is a promising step in this direction (Zhang et al. 2019). Without all this, data-driven science will be limited in its capabilities.

Computational materials science is presently dominated by the third research paradigm (cf. Fig. 1), but advancements in AI methods has been significant in recent years, and the fourth paradigm is playing an increasing role. Still, at present there is more hype than realism in what AI can do. Much of this relates to needs of domain-specific knowledge in materials science and engineering. Machine-learning techniques can already now help a lot when general trends are of interest and when one needs to fit and predict "the behavior of a big crowd" (see e.g. the methods used in the Kaggle competition for predicting properties of transparent conductors (Sutton et al. 2019). Often, the sensible needs of materials science and engineering are, however, different: We are typically not looking for a crowd behavior but we are searching for materials with extraordinary performance on certain functions or properties, often even a combination of several properties. There are typically just a few suitable materials among the enormous number of possible materials (already existing ones and those that will be synthesized in the future). However, in many cases we don't know how to identify this "class" of potentially interesting functional materials, and the number of *relevant* data contained in the used data sets is rather small as the few materials that we are searching for are statistically exceptional. How can we distinguish which data / materials are relevant and which are not? Learning about less than 0.01% relevant materials from thousands or millions of irrelevant data is obviously problematic, and standard methods, that optimize the regularized root-mean-square-error, even emphasize the importance of the irrelevant data, while surpassing the special cases. If the data could be grouped in "a majority classes" and a "minority class" methods have been developed to deal with the problem (Chawla et al. 2002 and references therein). However, often these classes are unknown and advancements of the subgroup discovery concept for the materials-science domain are urgently needed.

---

[11] See the chapter by M. Cohen in this handbook, in particular Fig. 4 (Cohen 2018).

What is missing at present? Let us list some issues:
- o Close coupling of materials property prediction with stability analysis and prediction of routes towards synthesis;
- o High-throughput studies of metastable materials and of the lifetimes of these metastable states;
- o Materials under real conditions ($T$, $p$, and reactive environment): stability and properties. This very much concerns multiscale modeling with robust, error-controlled links with knowledge of uncertainty between the various simulation methodologies. This has been often stated in the past but is still not fully realized;
- o Error estimates of calculations in terms of numerical approximations (basis sets, pseudopotentials, etc.) for specific properties (structure, elastic and electronic properties, etc.);
- o Computations beyond standard DFT as for example coupled-cluster methods for calculations for solids (possibly also getting prepared for quantum computers);
- o Complete description of scientific results accounting for the heterogeneity of data, i.e. to improve and complement present metadata definitions. While significant progress has been made toward transformation of computational data from the many computer codes and the development of corresponding metadata (Ghringhelli et al. 2016, 2017a, Meta Info at NOMAD) the advantage will only fully become apparent when the same will have been achieved also for experimental data. The latter challenge is even bigger than the first. The sample material used in the experimental study corresponds to the input file of a calculation; the experimental condition ($T$, $p$, environment) and the experimental equipment to the computer code. A not fully solved challenge is the definition of the sample materials. Obviously, closely coupled to the definition of metadata is the description of workflows in the sample preparation and running of the experiment.

The field is just developing the methods for the 4th paradigm. The learning curve connecting paradigms 1, 2, 3, and 4 is apparently getting steep. Thus the next paradigm may be close, even though the 4th has not been developed well, so far. What could be the next paradigm? Considering that "the future is already here – it's just not very evenly distributed" (Gibson 1999), it may hintingly visible already today. We guess that it may be virtual reality with direct and instantaneous connection to new calculations or a merger of theory (predictions and analysis) and experiment. There are exciting times ahead of us.

**Acknowledgements**

We gratefully acknowledge helpful discussions with Luca Ghiringhelli, Mario Boley, and Sergey Levchenko and their critically reading of the manuscript. This work received funding from the European Union's Horizon 2020 Research and Innovation Programme, Grant Agreement No. 676580, the NOMAD Laboratory CoE and No. 740233, ERC: TEC1P. We thank P. Wittenburg for clarification of the FAIR concept. The work profited from programs and discussions at the Institute of Pure and Applied Mathematics (IPAM) at UCLA, supported by the NFS, and from BIGmax, the Max Planck Society's Research Network on Big-Data-Driven Materials-Science.**References**